\documentclass[twocolumn,prl,aps,showpacs,floatfix,superscriptaddress]{revtex4}
\usepackage{graphicx}
\begin{document}

\widetext
\title{Most probable transition path in an overdamped system for a finite transition time}

\author{S.M. Soskin}
\altaffiliation{Also at Physics Department, Lancaster University,
UK}
\affiliation{Institute of Semiconductor Physics, National Academy of Sciences of Ukraine, 03028 Kiev, Ukraine}
\affiliation{The Abdus Salam ICTP, 34100 Trieste, Italy}

\begin{abstract}
The most probable transition path in a
one-dimensional overdamped system is rigorously proved to possess less than
two
turning points. The proof is valid for any potentials,
transition times, initial and final transition points.
\end{abstract}

\maketitle

\section{1. Introduction}

The overdamped stochastic equation is commonly defined as
\cite{FPE}

\begin{eqnarray}
&&
\frac{{\rm d}x}{{\rm d} t}=-\frac{{\rm d}U(x)}{{\rm
d}x}+f(t),
\\
&&
\langle f(t)\rangle =0, \quad \langle
f(t)f(0)\rangle =2D\delta(t).
\nonumber
\end{eqnarray}

The beginning of theoretical studies of such a  stochastic process
dates back to the celebrated works by Einstein
\cite{einstein} studying the Brownian motion of a free particle, which may
formally be considered as an overdamped stochastic motion in a parabolic potential
where the role of the generalized coordinate is played by the velocity.
A more general study of overdamped stochastic processes was started by
Smoluchowski \cite {smoluchowski} who formulated the equation of motion for
the probability density in an arbitrary overdamped system: this equation
bears his name nowadays. The next milestone was the work by Kramers \cite
{kramers} who, in particular, formulated the problem of the
noise-induced {\it escape} of an overdamped system from a metastable
potential well and derived its quasi-stationary solution:
the quasi-stationary
escape flux was found in \cite{kramers} using a stationary
solution of the Smoluchowski equation:

\begin{equation}
J_{qs}=A_{qs}{\rm e}^{-\frac{\Delta U}{D}}, \quad\quad D\ll\Delta
U,
\end{equation}

\noindent
where $\Delta U$ is the potential barrier (assumed to be much
less than the noise intensity $D$) and $A_{qs}$ is certain
prefactor which depends on $D$
weakly in comparison with the exponential (activation) factor.

As follows from \cite{kramers}, the escape flux becomes
quasi-stationary when time greatly exceeds a characteristic value
$t_{qe}\sim t_r\ln(\Delta U/D)$ where $t_r$ is a characteristic
relaxation time.
There were only a few theoretical works on the escape in overdamped
systems on time scales $t \lesssim t_{qe}$. One of the most general of such works was
the work by Shneidman \cite{schneidman} who solved the
non-stationary Smoluchowski equation for an arbitrary potential
using the method of the Laplace integral transformation while
assuming that the quasi-equilibrium in the vicinity of the bottom
of the well has been formed. The latter assumption is valid only
for times significantly exceeding the relaxation time $t_r$ while,
for shorter times, results of \cite{schneidman} are invalid.

The time scale $t \lesssim t_{r}$ was covered in the work
\cite{our} by means of the path-integral method \cite
{feinman,gy1960,alan1,lrh1} sometimes called also the method of
optimal fluctuation \cite
{lmd1998}. As a by-product, it was proved in \cite{our} that
the most probable escape path, i.e. the path providing the
absolute minimum of {\it action} in the functional space $[x(\tau)]$,
is monotonous i.e. $[x(\tau)]$ does not possess turning points.

In parallel to the development of the {\it escape} problem on short times,
there was an interesting discussion in the 90th
\cite{vugmeister,mannella,vugmeister1} on the {\it transition}
problem on short times.
This problem may be of interest in the context of the prehistory
probability density \cite
{prehistory} and of some biological problems \cite
{nigel,lindner}. Unlike the case of the escape, both the
initial and final points of the transition differ from the
stationary points of the noise-free system and, if they lie on one
and the same slope of the potential, the transition may possess
features distinctly different from those of the escape. Thus,
basing on the method of {\it optimal fluctuation} \cite {feinman,lmd1998},
the authors of
\cite{vugmeister} suggested that, for the short-time transition uphill
the slope of the potential barrier, the {\it most probable transition path} (MPTP)
may first relax close to the
bottom of the well and only then go to the final point. They
supported their suggestion by analytic calculations for the
parabolic approximation of the potential and, seemingly, by the
numerical calculations for the exact potential. However it was
shown in \cite{mannella} (also by means of numerical calculations
within the optimal fluctuation method) that the path
which first climbes up close to the barrier top and only then relaxes to
the final point may provide an exponentially larger activation
factor. Thus, just the latter path pretends to be the MPTP in such
a case.

The further development of this problem was done in \cite{our}: it
demonstrated that the {\it extreme} paths, i.e. paths providing
{\it local} minima of action, can possess {\it many} turning points;
\cite{our} provides the method how to explicitly calculate all
possible (for a given transition time) extreme paths and demonstrates that, as the
transition time increases, the MPTP may switch its topology from
the monotonous path to the path possessing one turning point,
either continuously or jump-wise.

The {\it present} work proves the general theorem stating that the
extreme paths possessing more than one turning point cannot
provide the {\it absolute} minimum of action i.e. the MPTP can be
only either monotonous or possessing just one turning point.

It should be noted also that,
apart from being necessary for a calculation of the activation
energy, the MPTP may be of interest on its own: e.g. in the
problem of the optimal control, the MPTP determines the
dynamics
of the external force which optimally enhances or suppresses a
given fluctuational transition \cite{sd,vr}.

\section{2. Basic equations}

In this section, I briefly reproduce basic equations of the
method of optimal fluctuation \cite{feinman,lmd1998} and those of
results \cite{our} which will be used in the next section for the
proof of the theorem.

Within the method of optimal
fluctuation, the flux is sought in the form

\begin{equation}
J(t)=P(D,t){\rm e}^{-\frac{S_a(t)}{D}}, \quad\quad D\ll S_a,
\end{equation}

\noindent
where the prefactor is assumed to depend on $D$ much more weakly than the
activation (exponential)
factor while the activation energy $S_a$ is a minimum of the
functional $S$, called action,

\begin{eqnarray}
&&
S_a=\min_{[x(\tau)]}S,\quad\quad
S\equiv S[x(\tau)]=\int_0^td\tau\; L,
\\
&&
L=\frac{1}{4}
\left(\dot{x}+
\frac{dU}{dx}
\right)^2,
\quad\quad
x(0)=x_0, \quad\quad x(t)=x_f.
\nonumber
\end{eqnarray}

The necessary condition for a minimum of the functional is the equality of the variation
$\Delta S$ to zero. The latter is equivalent to the {\it Euler equation} (EE):

\begin{equation}
\frac{\partial{L}}{\partial{x}}-
\frac{{\rm d}}{{\rm d}t}\left(\frac{\partial{L}}{\partial{\dot{x}}}\right)=0,
\end{equation}

\noindent
which, for a Lagrangian of the form (4), reads as

\begin{equation}
\ddot{x}+\frac{{\rm d}\tilde{U}}{{\rm d}x}=0, \quad\quad
\tilde{U}=-\frac{1}{2}\left(\frac{{\rm d}U}{{\rm d}x}\right)^2 .
\end{equation}

\noindent
So, solutions of the EE, called {\it extremal paths},
are
trajectories of the auxiliary mechanical
system (6).

The \textit{quasi-energy}

\begin{equation}
E\equiv -\partial S/\partial t=\dot{x}(\partial L/\partial \dot{x}) - L
=\frac{\dot{x}^2- (dU/dx)^2}{4}
\end{equation}

\noindent
is conserved along a solution of the EE, so that one easily derives from (7):

\begin{equation}
\dot{x}=\pm \sqrt{4E+(dU/dx)^2}.
\end{equation}

\noindent
It also
follows from (7) that the range of allowed quasi-energies is:

\begin{equation}
E\geq E_{min} \equiv -\min_{[x_0,x_f]} [ \left(dU/dx \right)^2/4
].
\end{equation}

Eq. (8) can be integrated in quadratures. Action $S$ can be expressed in
quadratures too.

For the case of {\it escape}, i.e. when the initial point is the
bottom of the well ($x_0=x_w$) \cite{stable_state}, $E_{min}=0$ and therefore the
motion in $\tilde{U}(x)$ with a quasi-energy $E\geq E_{min}$ cannot
possess turning points. Thus, the most probable escape path $[x(\tau)]$
is necessarily monotonous. On the contrary, in case of {\it
transition} within one slope of the potential i.e. when both $x_0$
and $x_f$ lie between the bottom of the well $x_w$ and the top of
the barrier $x_b$, the minimum quasi-energy $E_{min}$ is negative and
hence a trajectory of motion in the auxiliary potential $\tilde{U}(x)$
may possess a {\it negative} quasi-energy $E<0$. In the latter
case, the
trajectory possesses turning points in
$x_+$ and $x_-$,
which are the roots of
the equation

\begin{equation}
E= - \left(dU/dx \right)^2/4
\end{equation}

\noindent
where $x_{+/-}$ is the root nearest to $x_{f/0}$ among the
roots located at the same side of $x_{0/f}$ as $x_{f/0}$:

\begin{equation}
(x_{+/-}-x_{f/0})(x_{f/0}-x_{0/f})\geq 0.
\end{equation}

An extremal path for a given $E<0$ may turn in $x_-$ and $x_+$ any
number of times. Let us classify extremal paths by their topology,
namely by the overall number $N$ of turns of $[x(\tau)]$ (i.e. the number
of
changes of the sign of the velocity) and by the sign of the initial
velocity multiplied by the
sign of $x_f-x_0$: we shall use labels like \lq\lq$N=3,+$'' (in the case
of $N=0$, the ${\rm sign}[\dot{x}(x_f-x_0)]$ is necessarily \lq\lq$+$'',
so we shall omit the sign in the label in this case).
For each topology defined as above, the extremal path is uniquely
defined and, moreover, it can be implicitly expressed by means of
quadratures \cite{our}. The {\it full time} along an extremal of a given
topology can be {\it explicitly} expressed via quadratures.
To present these expressions in a compact form, it is convenient to introduce
three auxiliary ``times'':

\begin{eqnarray}
&&
t_0\equiv t_0(E)= t_{x_0\leftrightarrow x_f},\quad\quad
\nonumber
\\
&&
t_+\equiv t_+(E)=t_{x_f\leftrightarrow x_+},
\quad\quad
\nonumber
\\
&&
t_-\equiv t_-(E)=t_{x_-\leftrightarrow x_0},
\nonumber
\\
&&
t_{a\leftrightarrow b}\equiv
\left|\int_{a}^{b}\frac{dq}{\dot{q}(E,q)}\right|
={\rm sign}\left[\frac{b-a}{x_f-x_0}\right]\int_{a}^{b}dq\; z(q,E),
\nonumber
\\
&&
z(q,E)\equiv{\rm sign}[x_f-x_0]\frac{1}{\sqrt{4E+(dU(q)/dq)^2}}.
\end{eqnarray}

For different topologies, the dependence of the full time
along the extremal path on quasi-energy
reads as:

\begin{eqnarray}
&&
t_{N=0}(E)=t_0,
\\
&&
t_{N=2n+1,+/-}(E)=t_0+2t_{+/-}+(N-1)(t_0+t_{+}+t_-),
\nonumber
\\
&&
t_{N=2n+2,+/-}(E)=\pm t_0+N(t_0+t_{+}+t_-),
\nonumber
\\
&&
n=0,1,2,...
\nonumber
\end{eqnarray}

Figs. 1(c) and 2(c) show branches in the given ranges of $t$ and $E$, calculated by
(13),
for two characteristic cases related to Figs. 1(a,b) and 2(a,b)
respectively: when $-\tilde{U}(x)$ does not possess a local minimum
in between $x_w$ and $x_b$ (Fig. 1) and when it does (Fig. 2).

\begin{figure}[tb]
\includegraphics[width=1.6in,height=1.4in]{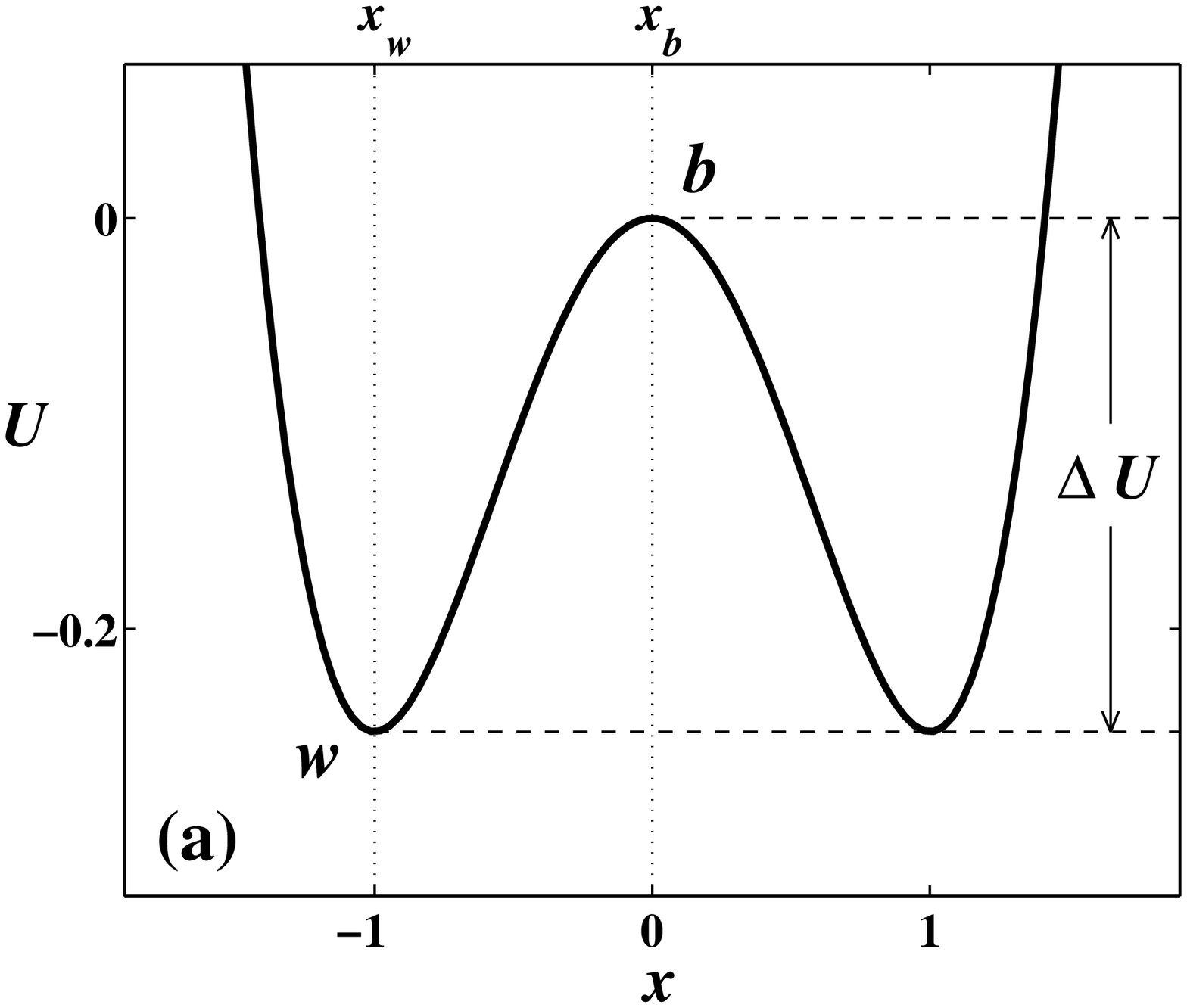}
\includegraphics[width=1.6in,height=1.4in]{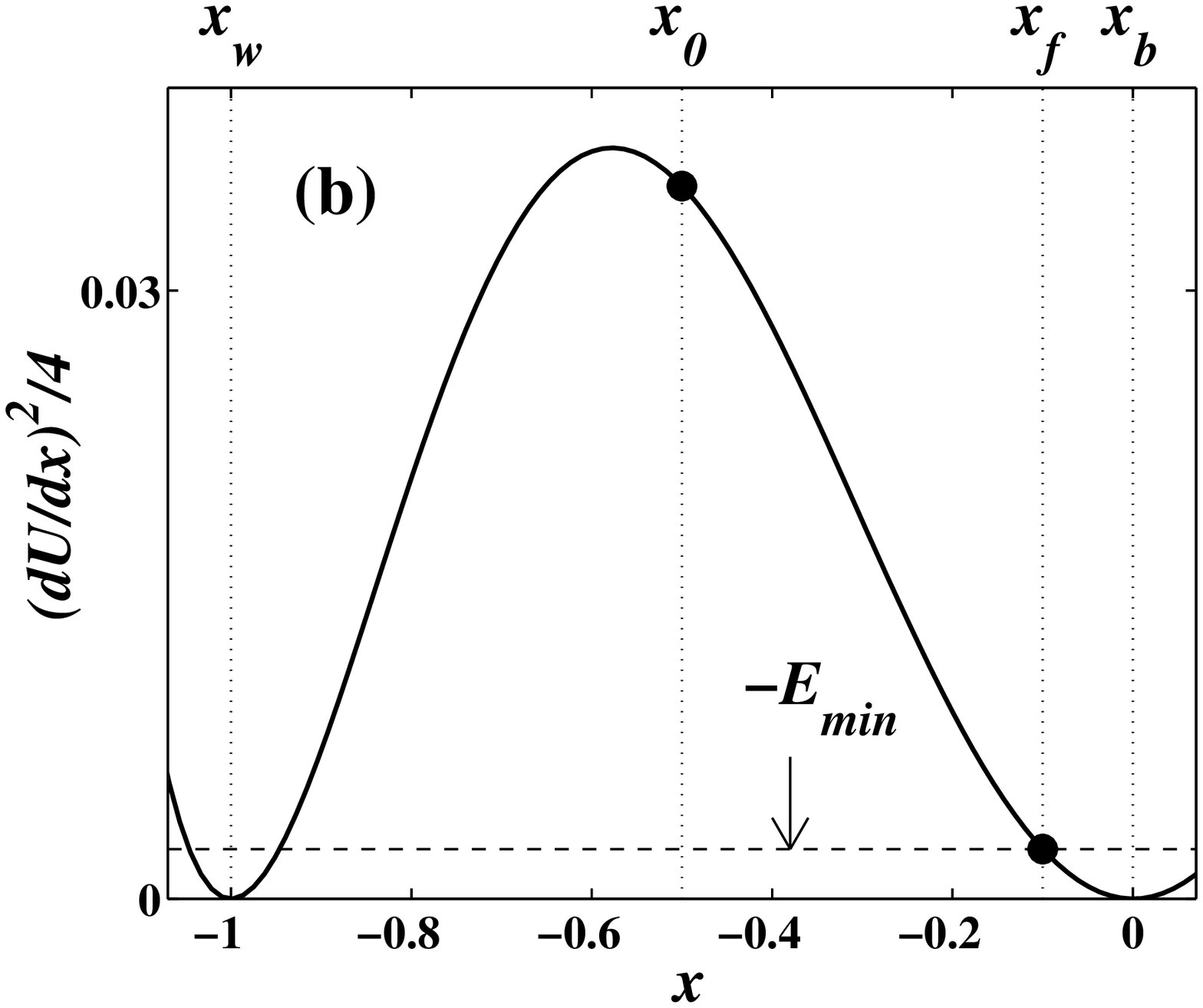}
\includegraphics[width=1.6in,height=1.4in]{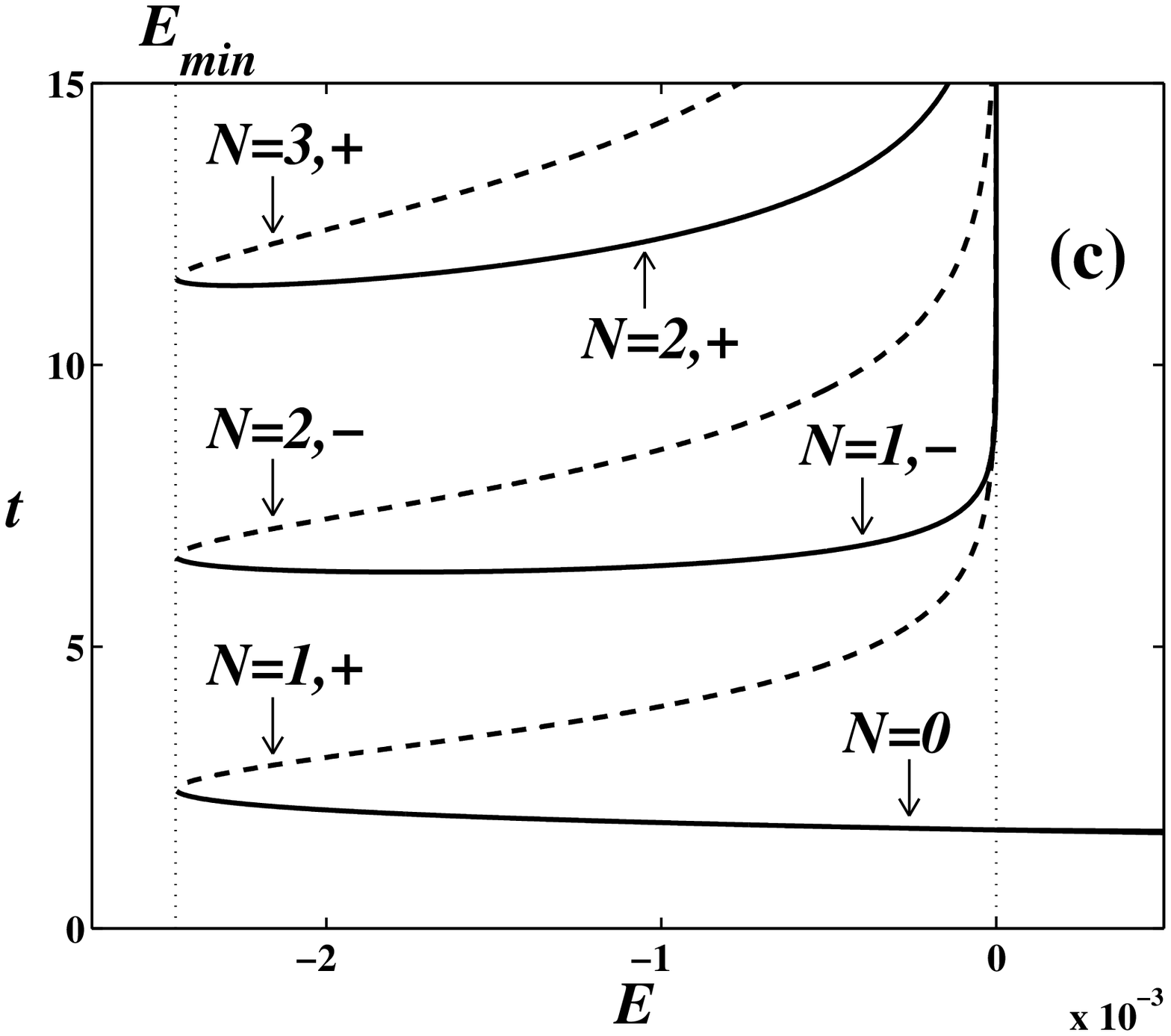}
\includegraphics[width=1.6in,height=1.4in]{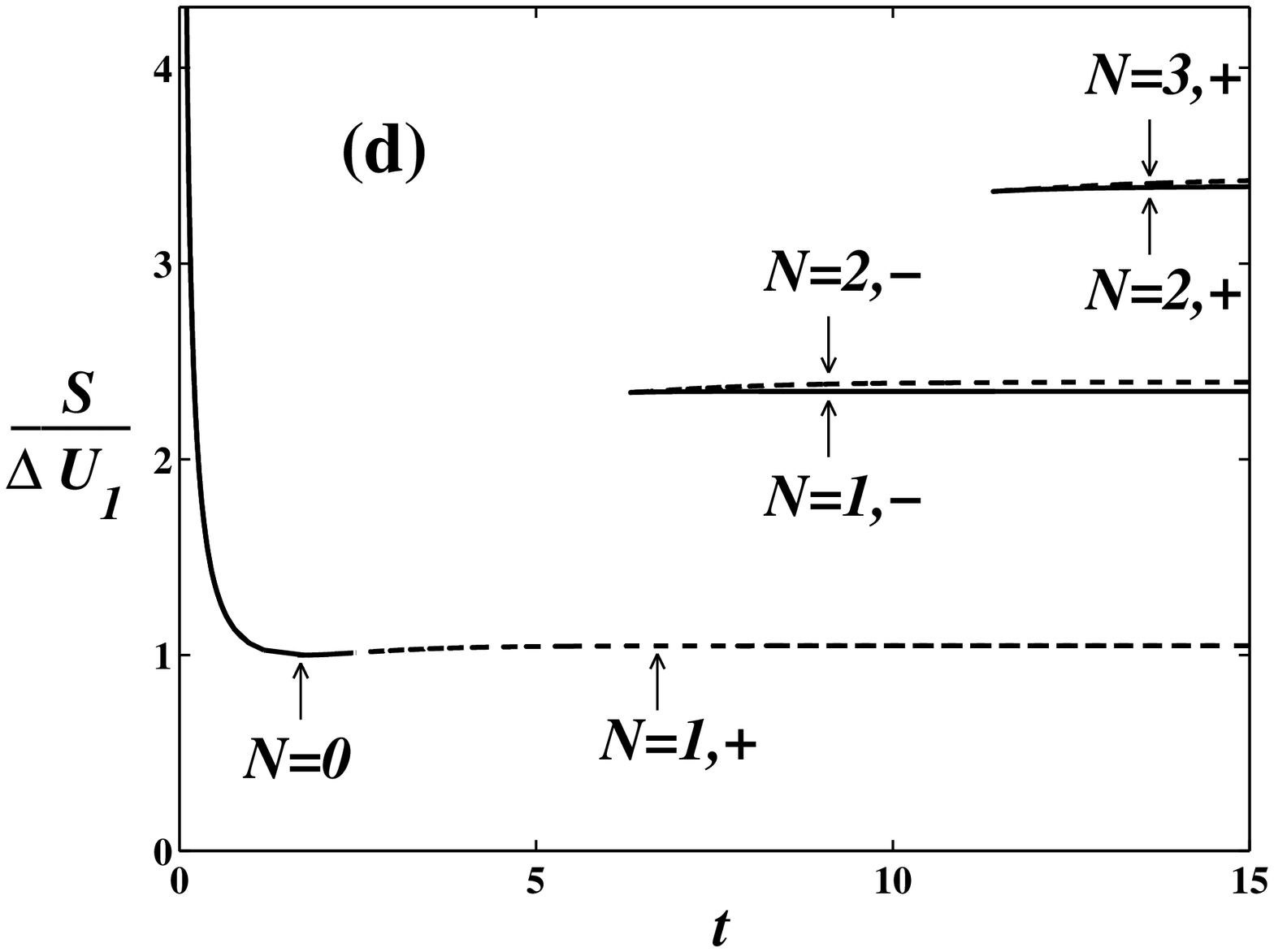}
\caption
{(a)The Duffing potential $U(x)=-x^2/2+x^4/4$; the left well
and the barrier are marked by the labels {\it w} and {\it b}
respectively (the cordinate $x_w$ of
the bottom of the well and the coordinate $x_b$ of the top of the barrier are indicated
by the dotted lines);
the potential barrier $\Delta U$ is indicated by the dashed lines.
(b) The function $(dU(x)/dx)^2/4\equiv -\tilde{U}(x)/2$
(thick solid line); the dots
show the points on the curve which
correspond to the initial and final points of the transition ($x_0=-0.5$ and
$x_f=-0.1$ respectively); $-E_{min}$ (9)
is indicated by the dashed line.
(c) Different branches of $t(E)$, calculated by Eq. (13) and
corresponding to different topologies of
the extremal path, are shown by thick solid/dashed lines with the labels
indicating the number of turning points and the sign
of the initial velocity $\dot{x}(0)$ multiplied by the sign of
$x_f-x_0$.
(d) Different branches of the action $S(t)$ (normalized by $\Delta U_1\equiv U(x_f)-U(x_0)=0.1044$),
calculated by Eq. (15), are marked similarly to the corresponding branches of
$t(E)$ in (c).
}
\end{figure}

\begin{figure}[tb]
\includegraphics[width=1.6in,height=1.4in]{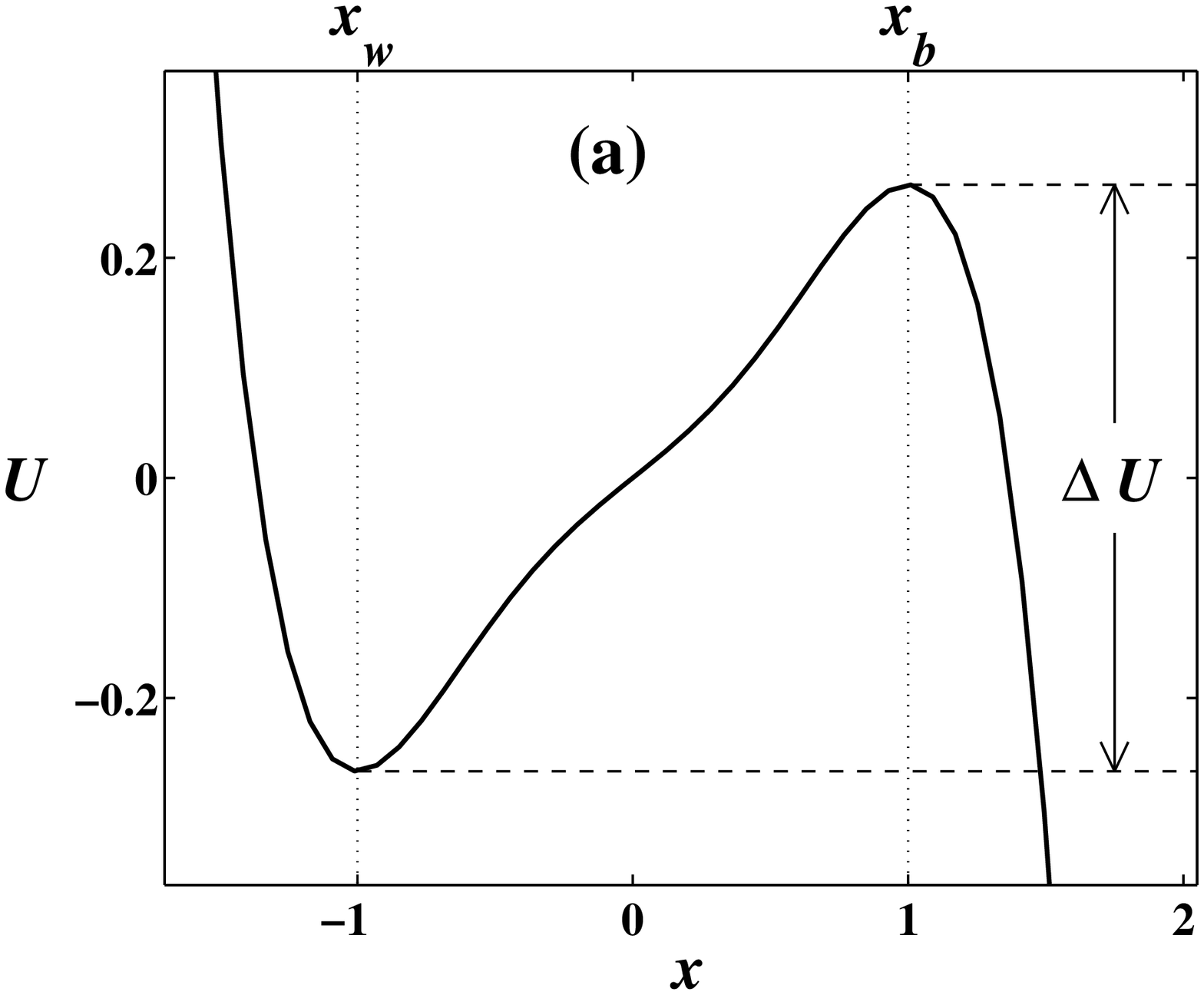}
\includegraphics[width=1.6in,height=1.4in]{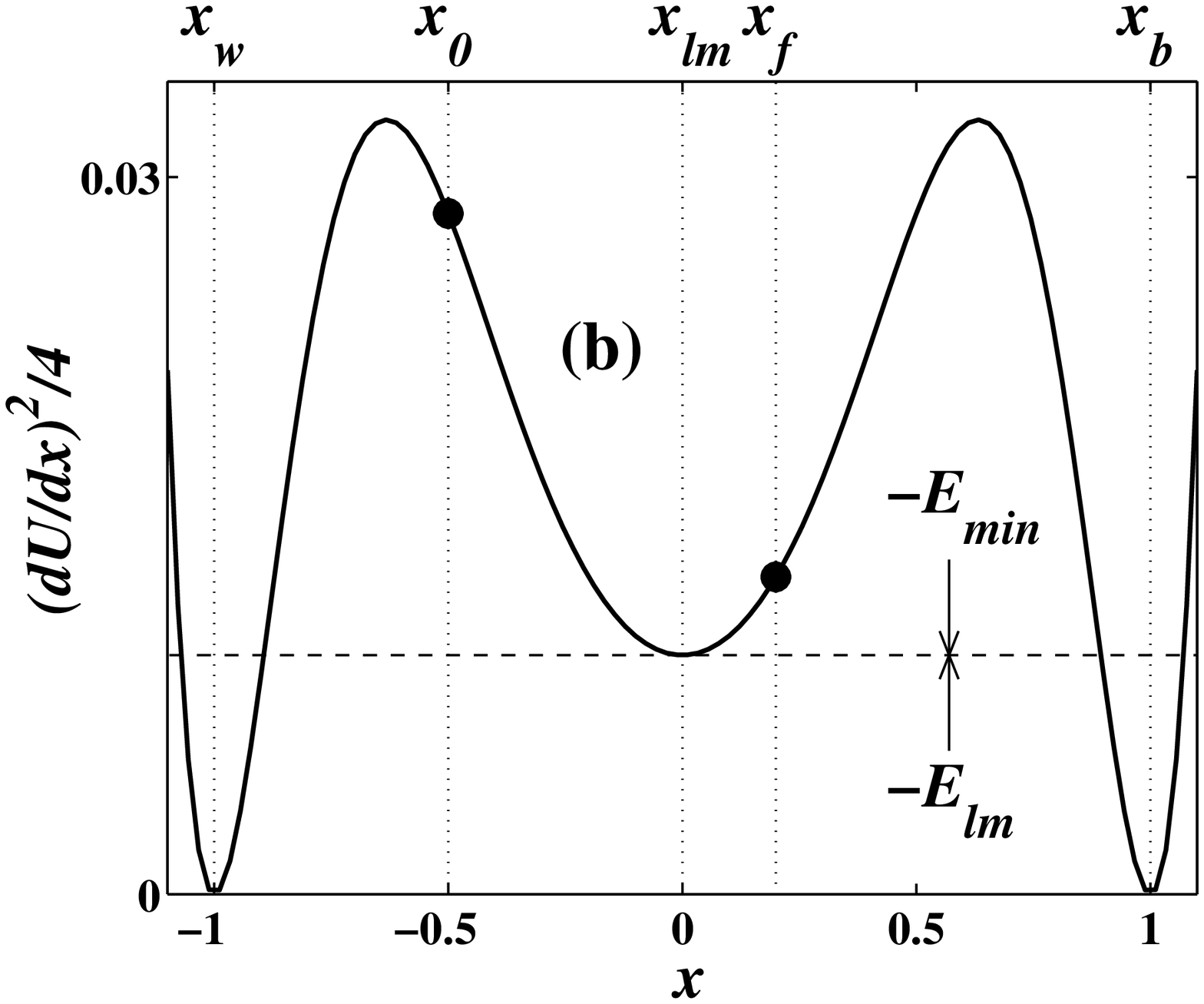}
\includegraphics[width=1.6in,height=1.4in]{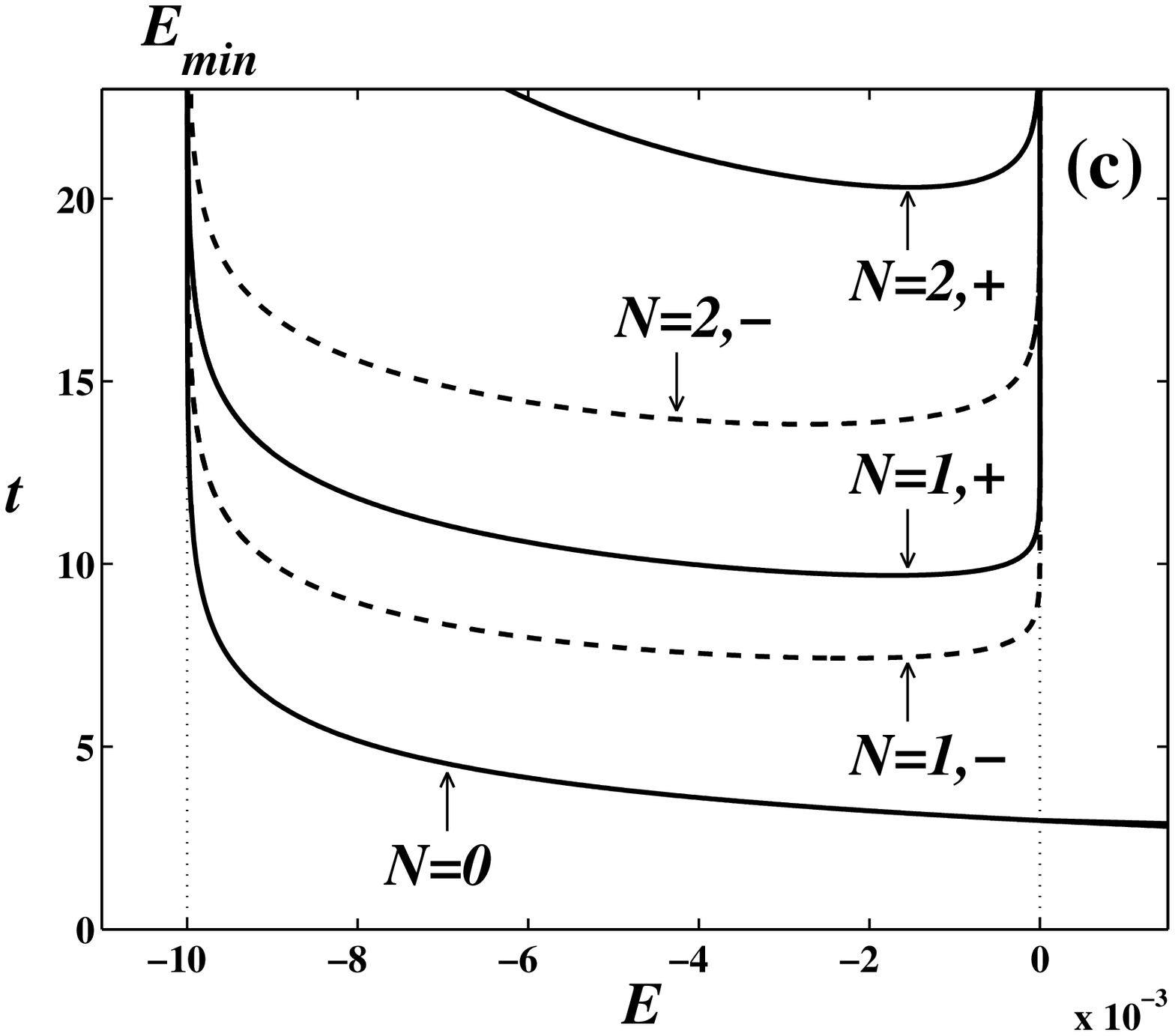}
\includegraphics[width=1.6in,height=1.4in]{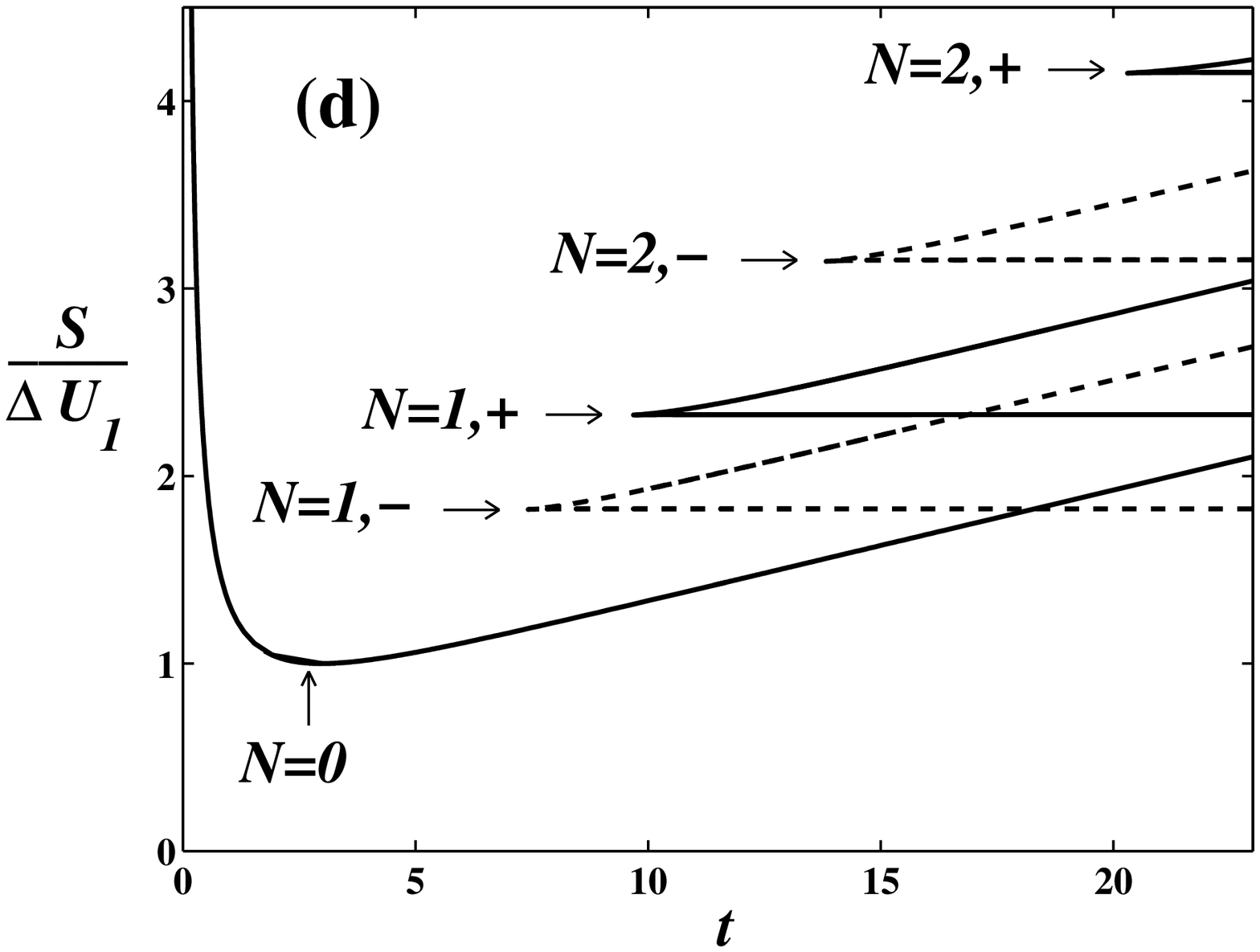}
\caption
{(a) The potential $U(x)=-x^5/5+0.8x^3/3+0.2x$; and (b) the
corresponding function $(dU(x)/dx)^2/4\equiv -\tilde{U}(x)/2$
(dots indicate points on the curve for
$x_0=-0.5$ and $x_f=0.2$); $E_{min}$
coincides with the singularity energy $E_{lm}$ related to
the local minimum of $(dU(x)/dx)^2/4$. Figures (c) and (d) are analogous to
Figs. 1(c) and 1(d) respectively. The normalization in (d) is:
$\Delta U_1\equiv U(x_f)-U(x_0)\approx 0.16915$.
}
\end{figure}

In order to present the
results for action in a compact form, it is convenient to introduce the auxiliary
actions:

\begin{eqnarray}
&&
S_0\equiv S_0(E)=\int_{x_0}^{x_f}dq\; \eta(q,E),
\\
&&
S_1\equiv S_1(E)=\int_{x_0}^{x_f}dq\; (\eta(q,E)-\frac{1}{2}dU/dq)
\nonumber
\\
&&
=S_0-\frac{1}{
2}\Delta U_1,
\quad\quad \Delta U_1\equiv U(x_f)-U(x_0),
\nonumber
\\
&&
S_+\equiv S_+(E)=\int_{x_f}^{x_+}dq\; (\eta(q,E)-\frac{1}{2}dU/dq),
\nonumber
\\
&&
S_-\equiv S_-(E)=\int_{x_-}^{x_0}dq\; (\eta(q,E)-\frac{1}{2}dU/dq),
\nonumber
\\
&&
\eta(q,E)=\frac{1}{2}
\left(
\frac{{\rm sign}[x_f-x_0](2E+(dU/dq)^2)}{\sqrt{4E+(dU(q)/dq)^2}}+
\frac{dU}{dq}
\right).
\nonumber
\end{eqnarray}

\noindent
Then $S(t)$ for various
branches can be shown to be as follows \cite{our}:

\begin{eqnarray}
&&
S_{N=0}(t)=S_0,
\\
&&
S_{N=2n+1,+/-}(t)=S_0+2S_{+/-}+(N-1)(S_1+S_{+}+S_-),
\nonumber
\\
&&
S_{N=2n+2,+/-}(t)=S_0+(\pm 1-1) S_1+N(S_1+S_{+}+S_-),
\nonumber
\\
&&
n=0,1,2,...,
\nonumber
\end{eqnarray}

\noindent
where $E\equiv E(t)$ in $S_0,S_1,S_+,S_-$ should be taken, for
a given branch $S_{N,+/-}(t)$,
as a solution of the equation

\begin{equation}
t=t_{N,+/-}(E),
\end{equation}

\noindent
where the functions $t_{N,+/-}(E)$ are defined
in (13) \cite{footnote3}.

Eqs. (12)-(16) describe in
quadratures all possible extremals and actions along them, in the
{\it general} case.

\section {3. Maximum number of turning points in the MPTP }


For Figs. 1 and 2, the
activation energy $S_a(t)$ (i.e. the minimal action) appears to be
constituted only by branches with
\lq\lq$N\leq 1$'' at any $t$. For the case like in Fig. 1, such a result is intuitively
predictible. But for the case like in Fig. 2, it is not so. Consider e.g. the case when
$x_0$ and $x_f$
are situated
in a relatively flat part of a potential while the potential
beyond it is much steeper (Fig. 3(a)). Intuition
might suggest that, if the transition time is large, then
multiple passages within the flat part of the potential might lead to
a smaller action than that for a path of the same duration but
with only one turning point: the latter path
might seem to necessarily involve one of the steep parts of the potential,
with a very large variation of the potential, which would lead in turn to a very large action.
So, the question arises whether it is a general property for the
number of turning points in the MPTP to be less than 2. We prove
below the theorem stating that it is.
As for the intuitive argument discussed above in relation to Fig. 3, it does not
contradict this theorem. Indeed, the MPTP does possess
less than two turning points while it still remains within the flat
part of the potential: it stays a main part of the given time in the minima of
$(dU/dx)^2$.

\begin{figure}[tb]
\includegraphics[width=1.6in,height=1.4in]{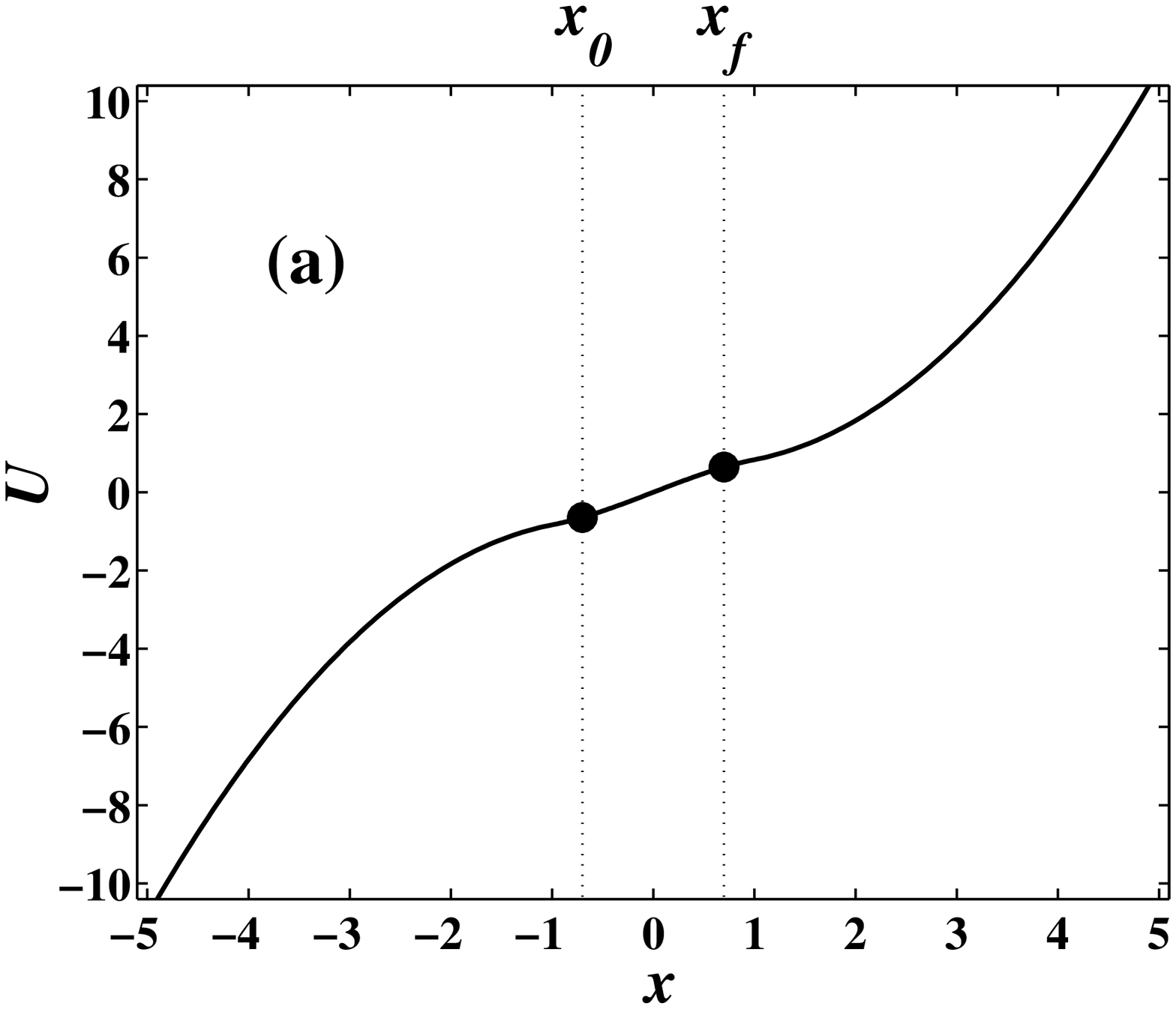}
\includegraphics[width=1.6in,height=1.4in]{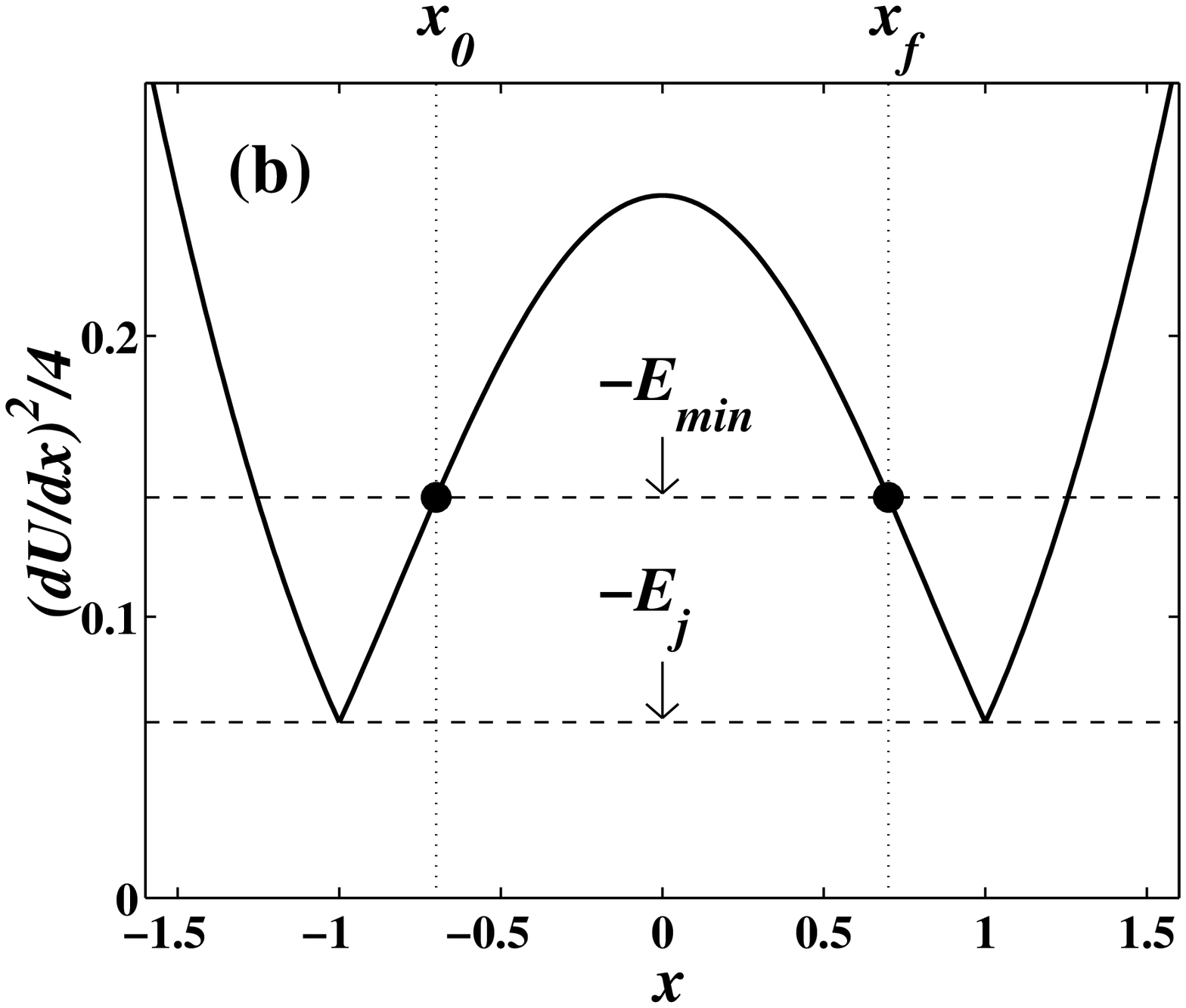}
\includegraphics[width=1.6in,height=1.4in]{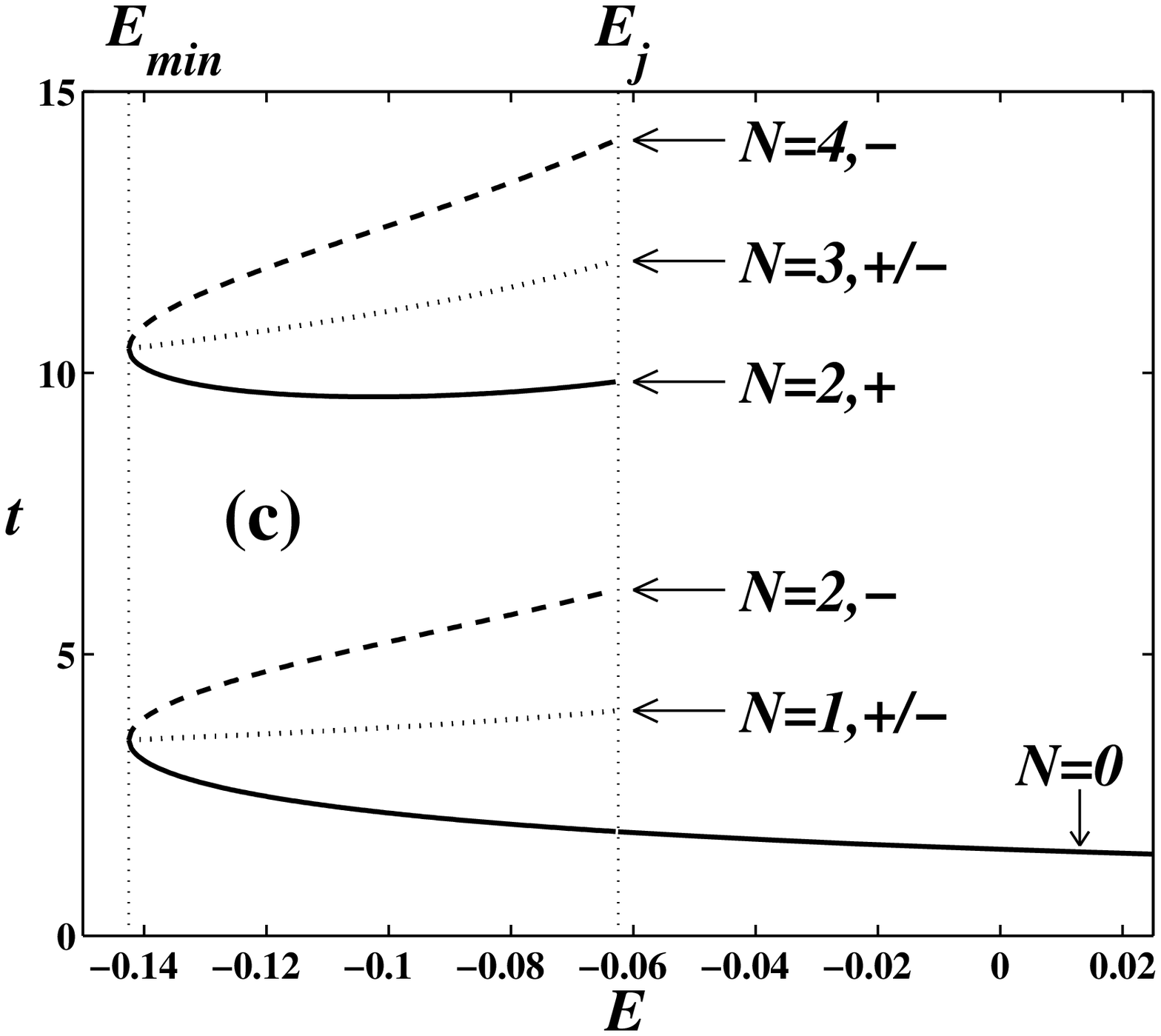}
\includegraphics[width=1.6in,height=1.4in]{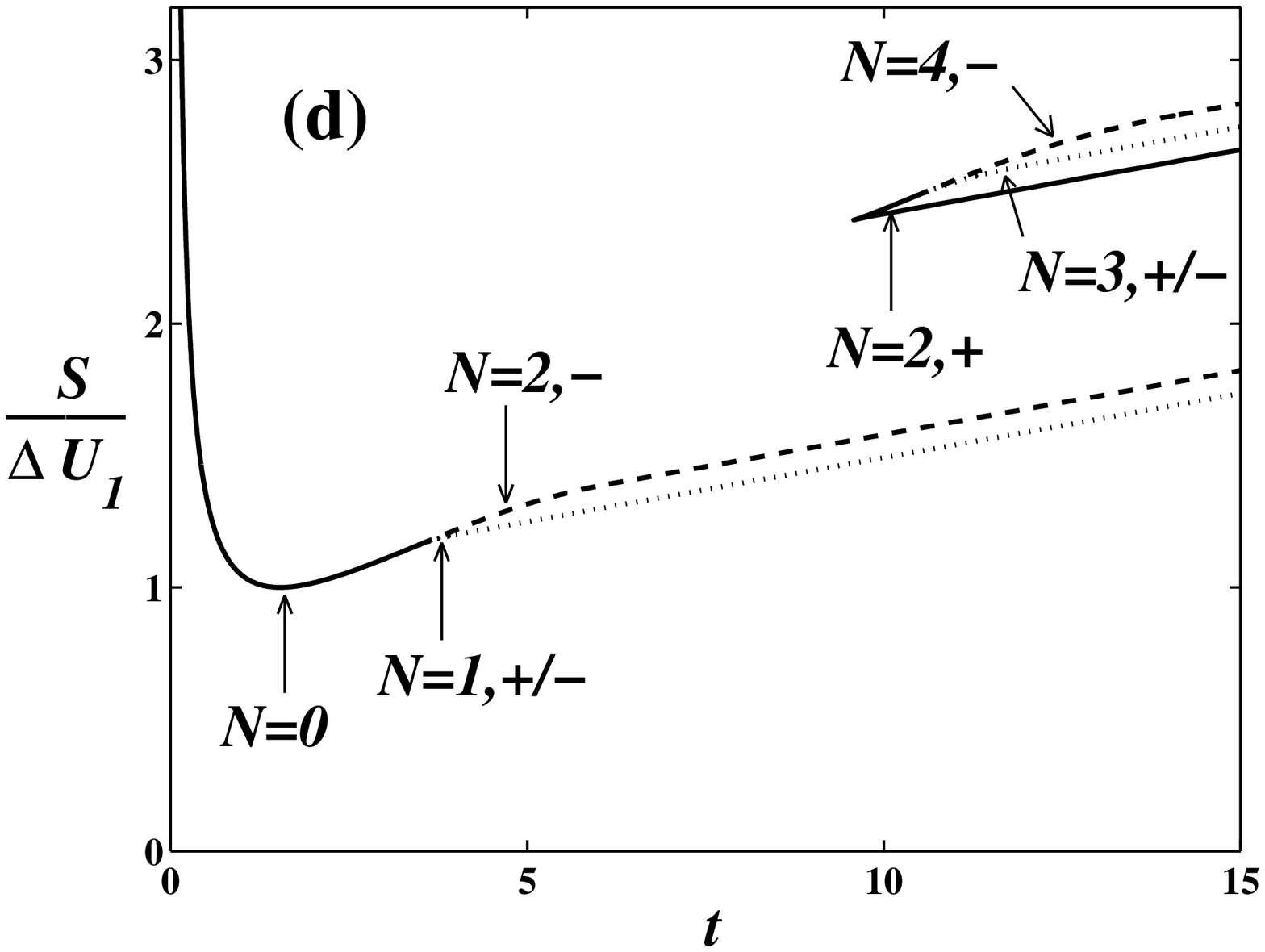}
\caption
{(a) The monotonously increasing potential $U(x)$ with the
following derivative: $dU/dx=1-0.5x^2$
at $x\in[-1,1]$ while $dU(x)/dx=0.5\mp(x\pm 1)$ at $\mp
x\in[1,\infty]$;
the initial and final transition points are indicated by dots and
and dotted lines:
$x_f=-x_0=0.7$.
(b) The function $({\rm d}U(x)/{\rm d}x)^2/4$ (thick solid line);
the level where ${\rm d}^2U/{\rm d}x^2$ undergoes the jump is
indicated by a dashed line and the label $-E_j$.
In (c) and (d), the branches
\lq\lq$N=0$'' and \lq\lq$N=2n,+$'' with $n=1,2,3,...$ are shown by the thick
solid lines while branches \lq\lq$N=2n,-$'' and \lq\lq$N=2n+1,+/-$''
are shown by dashed and dotted lines respectively. In (c), $\Delta
U_1\equiv U(x_f)-U(x_0)\approx 1.285667$. In (d),
$S(t)$
is calculated numerically by Eq. (15) in those ranges of $t$ where
the continuous paths exist while $S(t)$ is given in other ranges of $t$ by
Eq. (24) with $E=E_{j}$.
}
\end{figure}

\begin{figure}[tb]
\includegraphics[width=1.6in,height=1.4in]{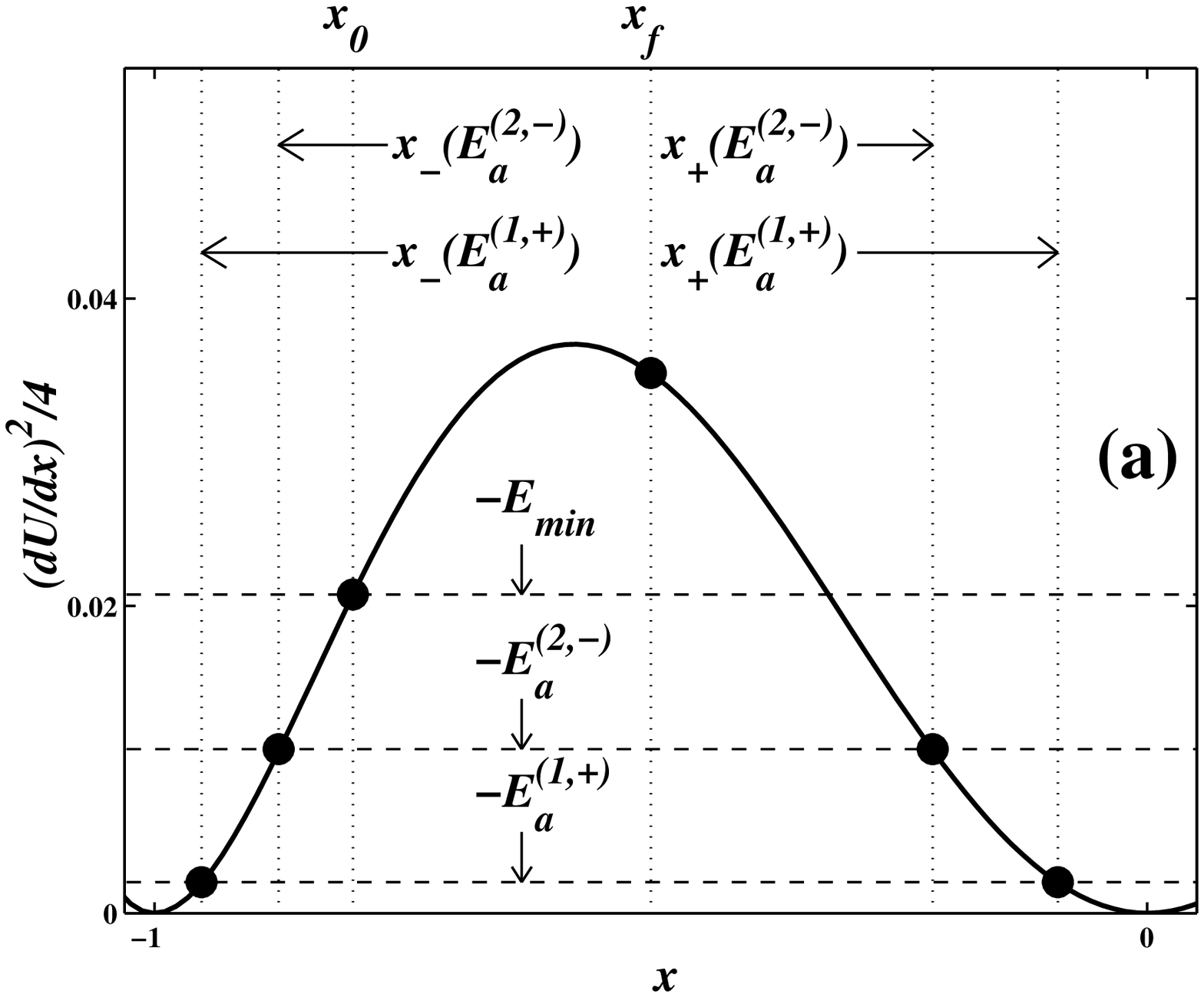}
\includegraphics[width=1.6in,height=1.4in]{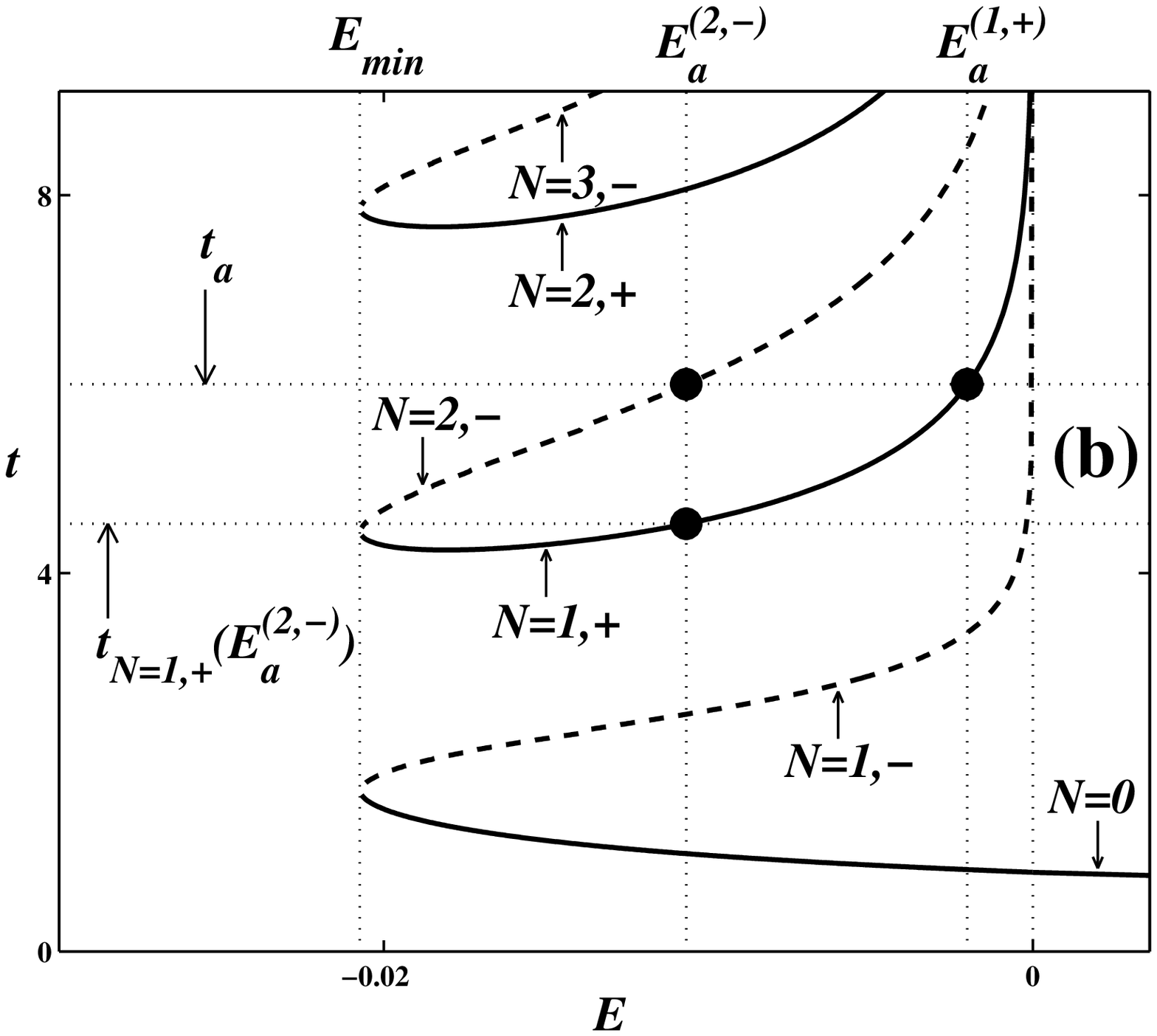}
\caption
{This figure illustrates the proof of the
relation $S_{N=2,-}(t)>S_{N=1,+}(t)$ at any $t\neq t_{N=2,-}(E_{min})\equiv t_{N=1,+}(E_{min})$.
For the sake of concreteness, we exploit the example of the
Duffing potential as in Fig. 1 while $x_0=-0.8$ and $x_f=-0.5$.
The relevant points are
marked by the large dots as well as indicated by thin dotted/dashed lines and by corresponding labels.
}
\end{figure}

\begin{figure}[tb]
\includegraphics[width=1.6in,height=1.5in]{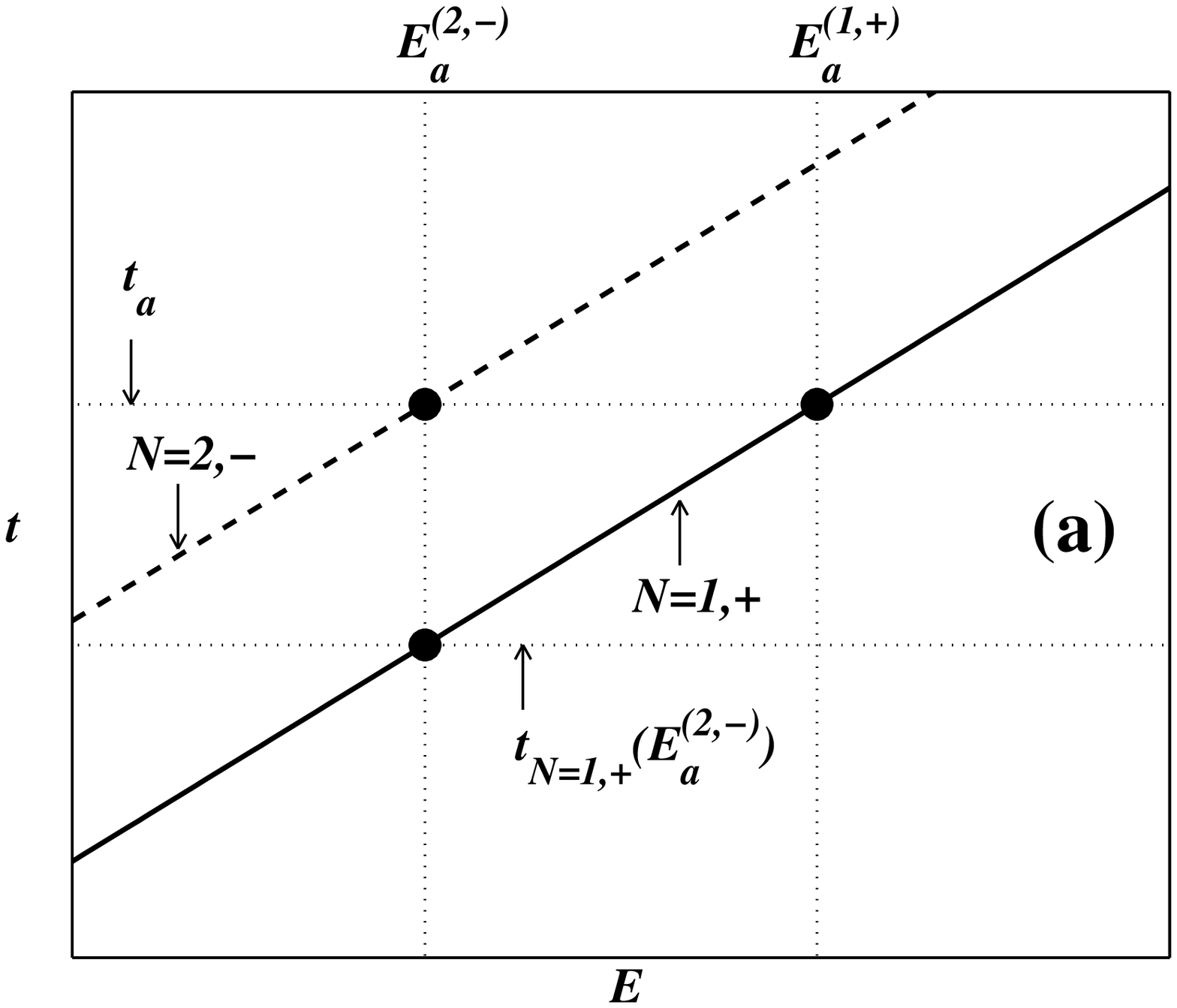}
\includegraphics[width=1.6in,height=1.5in]{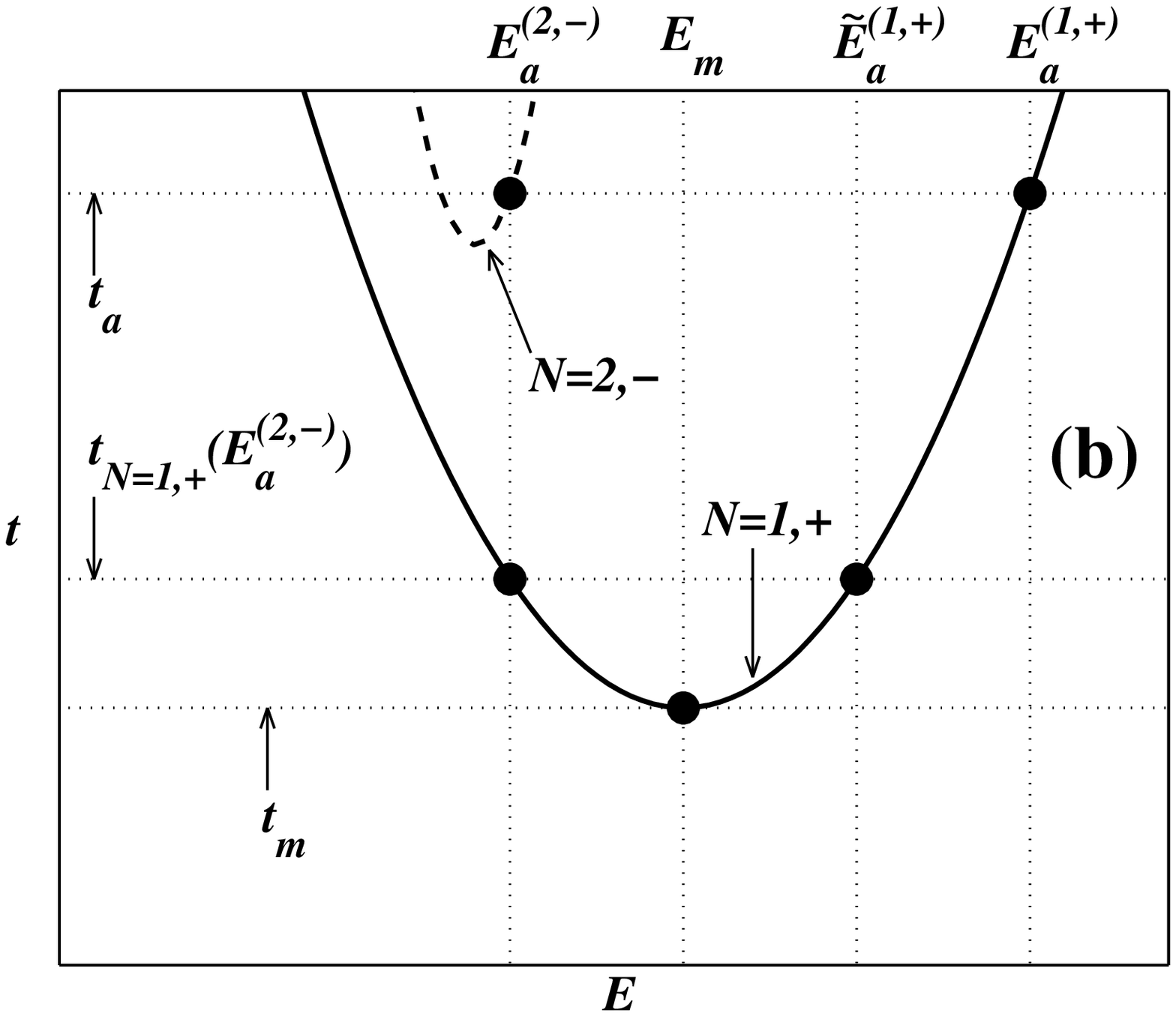}
\includegraphics[width=1.6in,height=1.5in]{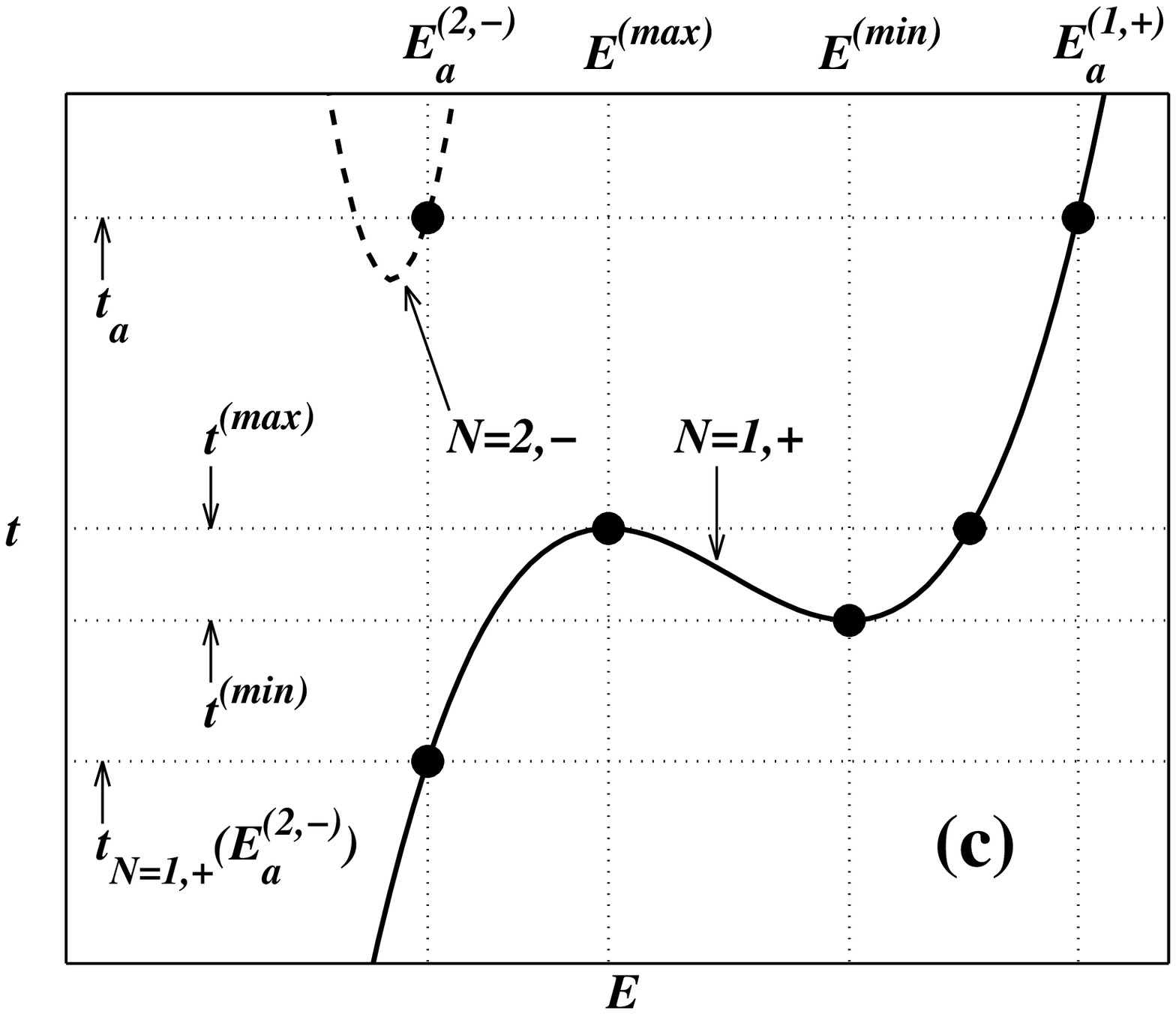}
\caption
{This figure shows schematically three characteristic cases
when the function $t_{N=1,+}(E)$ in the relevant range of energies
is (a) monotonous; (b) non-monotonous while there is no local maxima, (c) non-monotonous while there is
a local maximum.
The relevant points on the branches \lq\lq$N=2,-$'' and \lq\lq$N=1,+$'' are
marked by the large dots as well as indicated by thin dotted lines and by corresponding labels.
}
\end{figure}

{\bf Theorem:}
{\it the activation energy $S_a(t)$ is constituted
by the branches of action (15) with $N\leq 1$}.

{\bf Proof.}

1. Consider first the most common case, when {\it $dU(x)/dx$ is
continuous while $d^2U/dx^2$ is either continuous or, if it does
change jump-wise, is possessing one and the same
sign at both sides of the jump}.

For the sake of brevity, we assume below that $(dU(x_0)/dx_0)^2\leq (dU(x_f)/dx_f)^2$.
The case when the latter inequality does not hold can be proved analogously.

We use as an illustration the case shown in Fig. 4(a). Let us prove that

\begin{eqnarray}
&&
S_{N=2,-}(t)>S_{N=1,+}(t),
\\
&&
t\neq t_{N=2,-}(E_{min})\equiv t_{N=1,+}(E_{min})
\nonumber
\end{eqnarray}

\noindent
(at $t=t_{N=2,-}(E_{min})\equiv t_{N=1,+}(E_{min})$, the branches \lq\lq$N=2,-$'' and  \lq\lq$N=1,+$'' merge, so that the
actions obviously
coincide at this instant \cite{footnote_new}). Consider any time $t_a$ arbitrarily chosen from the
range where the equation

\begin{equation}
t_a=t_{N=2,-}(E)
\end{equation}

\noindent
possesses at least one root \cite{footnote4} (note that if the roots do exist they are necessarily
negative).
Consider any of the
roots of Eq. (18), $E_a^{(2,-)}$ (see Fig. 4(b)). As follows from
Eq. (13), the time $t_a$ can be presented in the following form:

\begin{eqnarray}
&&
t_a\equiv
t_{N=2,-}(E_a^{(2,-)})=
\\
&&
\quad
\quad
t_0(E_a^{(2,-)})+2(t_{+}(E_a^{(2,-)})+t_-(E_a^{(2,-)})),
\nonumber
\end{eqnarray}

\noindent
where $t_0,t_+,t_-$ are given in Eq. (12). As follows from Eq.
(15), the corresponding action can be presented as:

\begin{eqnarray}
&&
S_{N=2,-}(E_a^{(2,-)})=
\\
&&
S_0(E_a^{(2,-)})+2(S_{+}(E_a^{(2,-)})+S_-(E_a^{(2,-)})),
\nonumber
\end{eqnarray}

\noindent
where $S_0,S_+,S_-$ are given in Eq. (14).

Let us turn now to the branch \lq\lq$N=1,+$''.
As follows from (15),

\begin{equation}
S_{N=1,+}(E_a^{(2,-)})=S_0(E_a^{(2,-)})+2S_{+}(E_a^{(2,-)}).
\end{equation}

\noindent
Comparing (20) and (21), we may present $S_{N=2,-}(t_a)$ as

\begin{eqnarray}
&&
S_{N=2,-}(t_a)\equiv
S_{N=2,-}(E_a^{(2,-)})=S_{N=1,+}(E_a^{(2,-)})+
\nonumber
\\
&&
\quad\quad\quad\quad\quad\quad\quad\quad\quad\quad\quad\quad\quad\quad
\Delta S_{2,-}
\quad ,
\nonumber
\\
&&
\Delta S_{2,-}\equiv 2S_-(E_a^{(2,-)}).
\end{eqnarray}

Let us consider now the equation

\begin{equation}
t_a=t_{N=1,+}(E).
\end{equation}

\noindent
This equation necessarily possesses at least one (negative) root
which is larger than $E_a^{(2,-)}$. The latter is
a consequence of two properties: (i) $t_{N=1,+}(E)<t_{N=2,-}(E)$
since, at a given energy, the path \lq\lq$N=1,+$'' is a part of
the path  \lq\lq$N=2,-$''; (ii) $t_{N=1,+}(E)$ continuously
increases to $\infty$ as $E\rightarrow -0$ and as $E\rightarrow E_{lm}-0$ (the latter is relevant only to the case
with a local minimum of $(dU(x)/dx)^2$) \cite{footnote5}.
Fig. 5 illustrates this important property of an existence of a
root of Eq. (23) exceeding $E_a^{(2,-)}$.

Let us consider separately three characteristic cases
shown in Fig. 5. In all other cases, the proof can be reduced to the combination of those for these three
cases. Consider first the case when
$t_{N=1,+}(E)$ is monotonously increasing in the range $[E_a^{(2,-)},E_a^{(1,+)}]$
where $E_a^{(1,+)}$ is such a root of Eq. (23) which exceeds $E_a^{(2,-)}$ while being closer to
$E_a^{(2,-)}$ than any other root of Eq. (23) exceeding $E_a^{(2,-)}$
(see Fig. 5(a) or Fig. 4(b)). Using the property

\begin{equation}
 \frac{dS_{N,+/-}}{dt}=-E,
\end{equation}

\noindent
where $E\equiv E(t)$ is a solution of Eq. (16) \cite{24}, one may present $S_{N=1,+}(t_a)$
as

\begin{eqnarray}
&&
S_{N=1,+}(t_a)\equiv
S_{N=1,+}(E_a^{(1,+)})=S_{N=1,+}(E_a^{(2,-)})+
\nonumber
\\
&&
\quad\quad\quad\quad\quad\quad\quad\quad\quad\quad\quad\quad\quad\quad
\Delta S_{1,+}\quad,
\nonumber
\\
&&
\Delta S_{1,+}\equiv
\int_{t_{N=1,+}(E_a^{(2,-)})}^{t_a}dt \; (-E_{1,+}(t))\quad,
\end{eqnarray}

\noindent
where $E_{1,+}(t)<0$ is the function inverted towards the function
$t_{N=1,+}(E)$ in the relevant range of times $[t_{N=1,+}(E_a^{(2,-)}),t_a]$ and energies
$[E_a^{(2,-)},E_a^{(1,+)}]$.

As follows from (22) and (25), in order to prove (17), one needs
to prove

\begin{equation}
\Delta S_{2,-}>\Delta S_{1,+}\quad.
\end{equation}

\noindent
To do this, we shall estimate $\Delta S_{2,-}$ and $\Delta S_{1,+}$
from below and from above respectively and show that the estimate of $\Delta S_{2,-}$ from below provides,
at the same time,
the estimate of $\Delta S_{1,+}$ from above.

Let us first estimate $\Delta S_{2,-}$ from below:

\begin{eqnarray}
&&
\Delta S_{2,-}\equiv 2S_-(E_a^{(2,-)})
\nonumber
\\
&&
=\left |
 \int_{x_-(E_a^{(2,-)})}^{x_0}dq\;
 \frac{2E_a^{(2,-)}+(dU(q)/dq)^2}{\sqrt{4E_a^{(2,-)}+(dU(q)/dq)^2}}
\right |
\nonumber
\\
&&
\equiv
\left |
 \int_{x_-(E_a^{(2,-)})}^{x_0}dq\;
 \frac{-2E_a^{(2,-)}+[4E_a^{(2,-)}+(dU(q)/dq)^2]}{\sqrt{4E_a^{(2,-)}+(dU(q)/dq)^2}}
\right |
\nonumber
\\
&&
>\left |
 \int_{x_-(E_a^{(2,-)})}^{x_0}dq\;
 \frac{-2E_a^{(2,-)}}{\sqrt{4E_a^{(2,-)}+(dU(q)/dq)^2}}
\right |.
\end{eqnarray}

\noindent
The latter inequality is valid due to that both $-2E_a^{(2,-)}$ and $[4E_a^{(2,-)}+(dU(q)/dq)^2]$ are
necessarily positive.

Let us now turn to the estimate from above for $\Delta S_{1,+}$.
With this aim, we need to present $t_{N=1,+}(E_a^{(2,-)})$ in the following form (see Eqs. (13) and (19))

\begin{eqnarray}
&&
t_{N=1,+}(E_a^{(2,-)})=t_0(E_a^{(2,-)})+2t_{+}(E_a^{(2,-)})=
\nonumber
\\
&&
t_a-2t_-(E_a^{(2,-)}).
\end{eqnarray}

\noindent
The latter equality will be used in the last equality of Eq. (29):

\begin{eqnarray}
&&
\Delta S_{1,+}=\int_{t_{N=1,+}(E_a^{(2,-)})}^{t_a}dt \;
(-E_{1,+}(t))
\nonumber
\\
&&
\quad\quad\quad
<-E_a^{(2,-)}(t_a-t_{N=1,+}(E_a^{(2,-)}))=
\nonumber
\\
&&
\quad\quad\quad\quad
-2E_a^{(2,-)}t_{-}(E_a^{(2,-)}).
\end{eqnarray}

Allowing for Eq. (12) for $t_-$, one ultimately derives

\begin{equation}
\Delta S_{1,+}<
\left |
 \int_{x_-(E_a^{(2,-)})}^{x_0}dq\;
 \frac{-2E_a^{(2,-)}}{\sqrt{4E_a^{(2,-)}+(dU(q)/dq)^2}}
\right |.
\end{equation}

Comparing the inequalities (27) and (30), one immediately
derives (26), while
the latter together with (22) and (25) prove the inequality (17).

The case of the non-monotonous (in the range $[E_a^{(2,-)},E_a^{(1,+)}]$)
dependence of $t_{N=1,+}(E)$ (see Figs. 5(b) and 5(c)) may be treated
similarly. The only difference is that, in this
case, $E_{1,+}(t)$ is a
multi-valued function so that it is necessary to divide the range
$[E_a^{(2,-)},E_a^{(1,+)}]$
for ranges in each of which $E_{1,+}(t)$ is a
single-valued function.
Consider first the case when $t_{N=1,+}(E)$ possesses in the range $[E_a^{(2,-)},E_a^{(1,+)}]$
only a local minimum while not possessing local
maxima (Fig. 5(b)).
Denoting energy of the local minimum as $E_m$ and denoting $E_{1,+}(t)$ in the ranges
$[E_a^{(2,-)},E_m]$ and
$[E_m,E_a^{(1,+)}]$ as $E^{(1)}_{1,+}(t)$ and $E^{(2)}_{1,+}(t)$
respectively, one may present $\Delta S_{1,+}$ as

\begin{eqnarray}
&&
\Delta S_{1,+}=
\nonumber
\\
&&
\quad\quad\quad
\int_{t_{N=1,+}(E_a^{(2,-)})}^{t_m}dt \;
(-E^{(1)}_{1,+}(t))+\int_{t_m}^{t_a}dt \;
(-E^{(2)}_{1,+}(t))
\nonumber
\\
&&
\quad\quad\quad =\int^{t_{N=1,+}(E_a^{(2,-)})}_{t_m}dt \;
(E^{(1)}_{1,+}(t)-E^{(2)}_{1,+}(t))+
\nonumber
\\
&&
\quad\quad\quad\quad
\int_{t_{N=1,+}(E_a^{(2,-)})}^{t_a}dt \;
(-E^{(2)}_{1,+}(t)).
\end{eqnarray}

\noindent
The integrand in the first integral in the last right-hand side is
negative in the whole range $[t_m,t_{N=1,+}(E_a^{(2,-)})]$ and
therefore the integral is negative too. As for the second
integral, it can be estimated from above in the following way:

\begin{eqnarray}
&&
\int_{t_{N=1,+}(E_a^{(2,-)})}^{t_a}dt \;
(-E^{(2)}_{1,+}(t))<
\nonumber
\\
&&
(-\tilde{E}_a^{(1,+)})(t_a-t_{N=1,+}(E_a^{(2,-)}))<
\nonumber
\\
&&
(-E_a^{(2,-)})(t_a-t_{N=1,+}(E_a^{(2,-)}))
\end{eqnarray}

\noindent
(for the sake of clarity, the notation $\tilde{E}^{(1,+)}\equiv E_{1,+}^{(2)}(t_{N=1,+}(E_a^{(2,-)}))$
has been introduced in the middle line in Eq. (32); see also Fig. 5(b)).
The last right-hand side in (32) is exactly the same as in the middle
line in (29) so that its further estimate is identical to that in
(29)-(30). Given that the first integral in the last right-hand side
in (31) is negative, the inequality (26) is satisfied in this case
even stronger than in the case of a monotonous function $t_{N=1,+}(E)$.

Finally, let us briefly consider the case of $t_{N=1,+(E)}$
possessing a local maximum (Fig. 5(c)). Denoting energies of the
local maximum and minimum \cite{minimum} as $E^{(max)}$ and $E^{(min)}$ respectively,
and denoting $E_{1,+}(t)$ in the ranges
$[E_a^{(2,-)},E^{(max)}]$, $[E^{(max)},E^{(min)}]$ and
$[E^{(min)},E_a^{(1,+)}]$ as $E^{(1)}_{1,+}(t)$, $E^{(2)}_{1,+}(t)$ and $E^{(3)}_{1,+}(t)$
respectively, one may present $\Delta S_{1,+}$ as

\begin{eqnarray}
&&
\Delta S_{1,+}=
\int_{t_{N=1,+}(E_a^{(2,-)})}^{t^{(max)}}dt \;
(-E^{(1)}_{1,+}(t))+
\\
&&
\int_{t^{(max)}}^{t^{(min)}}dt \;
(-E^{(2)}_{1,+}(t))+\int_{t^{(min)}}^{t_a}dt \;
(-E^{(3)}_{1,+}(t))=
\nonumber
\\
&&
\int^{t^{(min)}}_{t^{(max)}}dt \;
(E^{(2)}_{1,+}(t)-E^{(3)}_{1,+}(t))+
\nonumber
\\
&&
\left[\int_{t_{N=1,+}(E_a^{(2,-)})}^{t^{(max)}}dt \;
(-E^{(1)}_{1,+}(t))+
\int_{t^{(max)}}^{t_a}dt \;
(-E^{(3)}_{1,+}(t))
\right].
\nonumber
\end{eqnarray}

\noindent
The integrand in the first integral in the last right-hand side is
negative in the whole range $[t^{(min)}, t^{(max)}]$ and therefore the
integral is negative too. As for the expression in the
brackets, it can be estimated from above in the following way:

\begin{eqnarray}
&&
[\int_{t_{N=1,+}(E_a^{(2,-)})}^{t^{(max)}}dt \;
(-E^{(1)}_{1,+}(t))+
\int_{t^{(max)}}^{t_a}dt \;
(-E^{(3)}_{1,+}(t))]
\nonumber
\\
&&
<-E_a^{(2,-)}(t_a-t_{N=1,+}(E_a^{(2,-)})),
\end{eqnarray}

\noindent
which is exactly the same as in (29) or (32), so that its
further estimate is the same too. Thus, the inequality (26) is
satisfied even stronger than in the monotonous case.

Thus, we have proved the inequality (17) both for monotonous and non-monotonous $t_{N=1,+}(E)$.
One can similarly
prove the inequality

\begin{equation}
S_{N=2,+}(t)>S_{N=1,-}(t),
\end{equation}

\noindent
and similar inequalities related to branches with $N>2$. So, the
theorem has been proved for the case 1.

2. Consider now the formal case when {\it
$d^2U/dx^2$ changes its sign jump-wise at some $x=x_j$}. Formally,
the branches $t(E)$ for the solutions of the Euler equation abrupt
at the discontinuity energy $E_j$ (Fig. 3(b)) so that there may be
time ranges where the Euler equation does not possess any solution
at all. However, it means only that there is no {\it continuous}
path which would provide a minimum action in this time range; the discontinuous path
does exist though. In order to obtain its continuous
approximation, it is necessary to approximate the jump of $d^2U/dx^2$
by any sequence of continuous functions whose relevant coordinate
range converges to the infinitesimal vicinity of the
coordinate of the jump, $x_j$. Then the continuous extremal paths
do exist while, as it follows from Eq. (24), the limit of the
sequence of the corresponding actions is equal to (cf. Fig. 3(c))

\begin{equation}
S(t)=S(t(E_j))- E_j(t-t(E_j)),\quad\quad t>t(E_j),
\end{equation}

\noindent
where $t(E_j)$ is an upper limit for a given branch in the discontinuous
case (cf. Fig. 3(b)). The limit of the sequence of the corresponding continuous
extremal paths is the following discontinuous path: its parts which correspond to $x$ beyond the immediate
vicinity of $x_j$ coincide with those of the corresponding continuous
path for $E=E_j$ while it stays in $x=x_j$ during the rest of the
time i.e. during the interval $t-t(E_j)$. Given that each path from the sequence satisfies the
theorem, so does their discontinuous limit.

3. The formal case when {\it $dU/dx$ is discontinuous} can be
proved analogously to the case 2 above.

Altogether, this proves the theorem on the whole.

\section {4. Conclusions}

In this Letter, I have rigorously proved that, in the problem of the noise-induced
transition between points lying within a monotonous part of a potential of an
overdamped one-dimensional system, action corresponding to
solutions of the Euler equation which possess two or more turning
points is necessarily larger than action corresponding to
solutions with one turning point. This means that the {\it most probable transition
path cannot possess more than one turning point}. The proof is
general, i.e. valid for any potential, any positions of the
initial and final transition points, and any value of the
transition time. The practical use of the theorem proved in the
Letter consists in the following: if one needs
to calculate the
non-stationary transition flux or probabilty density or any other
quantity related to the non-stationary transition caused by a weak noise,
then one may
skip the calculation and analysis of all (sometimes numerous) partial
contributions corresponding to solutions of the Euler equation
which possess more than one turning point.

\section{Acknowledgements}

I express my gratitude to T.L. Linnik for a
technical assistance at the drawing Figs. 3(c,d), 4(b). I also acknowledge the
discussion with V.I. Sheka as well as appreciate the stimulating
role of R. Mannella for me to write this paper.


\begin{thebibliography}{1}
\bibitem{stable_state} The bottom of the well is the stable
stationary state of the noise-free system.
\bibitem{footnote3} For the case of the Duffing potential and
the set of initial and final points as in Fig. 1, Eq. (16) has no
more than one root.
But, generally, Eq. (16) may have more than one root (cf.
Fig. 2(c)), so that the function $S_{N,+/-}(t)$
is multi-valued then (cf. Fig. 2(d)).
\bibitem{footnote_new}
For $t=t_{N=2,-}(E_{min})$, the second turning point of the path
\lq\lq$N=2,-$'' degenerates, so that the theorem obviously holds
for this $t$ too.
\bibitem{footnote4} Note that the branches $t_{N=0}(E)$ and $t_{N=1,+/-}(E)$
cover together the whole range of time $[0,\infty]$ as $E$ varies.
Thus, for those $t_a$ at which Eq. (18) does not possess any
root, $S_a(t)$ obviously cannot be constituted by the branch
\lq\lq$N=2,-$'' while it can be constituted by those with $N\leq
1$.
\bibitem{footnote5} It is right the property (ii) that requires the
absence of a jump-wise change of the sign of $d^2U(x)/dx^2$
and of a jump-wise turn of $d^2U(x)/dx^2$ into zero. Otherwise,
the time range covered by the branch \lq\lq$N=1,+$'' would be
limited from above: cf. Fig. 3.
\bibitem{24} Eq. (24) can be proved analogously to the similar equation for the
mechanical action \cite{Landau:76}.
\bibitem{minimum} Given the definition of
$E_a^{(1,+)}$ (see the paragraph preceding Eq. (24)),
the local minimum should necessarily exist if $t_{N=1,+}(E)$
is assumed non-monotonous in the range $[E_a^{(2,-)},E_a^{(1,+)}]$.
\end{thebibliography}
\end{document}